\documentstyle[12pt]{article}
\title{Gravity of Monopole and String and Gravitational Constant  in $^3$He-A.}
\author{G.E. Volovik\\
Low Temperature Laboratory,
Helsinki University of Technology\\
P.O.Box 2200, FIN-02015 HUT, Finland\\
and\\
L.D. Landau Institute for Theoretical Physics,
 Moscow\\
}
\begin{document}
\maketitle
\begin{abstract}
{ We discuss the effective metric produced in superfluid $^3$He-A by such
topological objects as radial disgyration and monopole. In relativistic
theories  these metrics are similar to that of the local string and global
monopole correspondingly. But in $^3$He-A they have the negative angle deficit,
which corresponds to the negative mass of the topological objects. The
effective
gravitational constant in superfluid $^3$He-A, derived from the comparison
with relativistic theories, is
$G\sim \Delta^{-2}$, where the gap amlitude $\Delta$ plays the part of the
Planck
energy cut-off. $G$ depends on temperature roughly as
$(1-T^2/T^2_c)^{-2}$, which corresponds to the vacuum screening of the Newton's
constant.
 }
\end{abstract}

 Submitted to JETP Letters. \\

{\bf Introduction.}
The effective gravity arises in many condensed matter systems. The
typical
examples are the crystal with dislocations and disclinations, which
models the
effective space with curvarture and torsion (see the References in
the latest
papers \cite{Katanaev,Baush} on this subject), and the normal (or
superfluid)
liquid where the sound waves (or phonons) propagate in the effective
Lorentzian
space generated by the background (super) flow
\cite{UnruhSonic,Visser1997}.
However it appears that the superfluid $^3$He-A provides the most
adequate
analogies for
the relativistic models of the effective gravity, which allow to
simulate many
different properties of the quantum vacuum
\cite{VolovikVachaspati,JacobsonVolovik,Exotic,AxialAnomaly}.

The quasiparticles in
$^3$He-A are chiral and massless fermions. Their spectrum is
determined by 3
parameters. One of them is the gap amplitude $\Delta(T)$ which plays
the
part of
the Planck cut-off energy. Below this cut-off, which depends on
temperature
$T$,
the fermions are "relativistic" with the spectrum:
\begin{equation}
E^2({\bf k}) + g^{ik}(k_i -eA_i) (k_k -eA_k)=0~~.
\label{E2}
\end{equation}
Here  ${\bf A}$ is the dynamical vector potential of the induced
"electromagnetic field",
${\bf A}=k_F{\hat{\bf l}}$, where ${\hat{\bf l}}$ is unit vector in
the
direction of the gap nodes  in the  momentum space. The same vector
determines
the uniaxial anisotropy of the metric tensor of the effective space
which
governs the motion of fermions. In equilibrium this metric is
\begin{equation}
g^{ik}=- c_\perp^2 (\delta^{ik} -   l^i  l^k) -  c_\parallel^2
l^i
 l^k ~~,~~g^{00}=1~.
\label{gik}
\end{equation}
Here $c_\perp= \Delta(T) / p_F$ and $c_\parallel=v_F$ (with $c_\perp
\ll c_\parallel $) are the "speeds of
light" propagating transverse to
${\hat l}$ and along ${\hat l}$ correspondingly. The Fermi momentum
$k_F$ and
the Fermi velocity  $v_F$ are practically independent from $T$ while
 $\Delta (T)$ strongly depends on $T$:  $\Delta^2(T)\sim
\Delta^2(0) (1-T^2/T_c^2)$, where $T_c \sim \Delta(0)$ is the
temperature of
superfluid transition.

If the
$\hat l$-field is homogeneous, say,
$\hat l=\hat z$, the anisotropy of the metric in Eq.(\ref{gik}) can be
removed by
rescaling:
$  z= c_\parallel Z$, $ x= c_\perp X$ and $ y= c_\perp Y$.  However if
$\hat l$-field is inhomogeneous, the effective metric gains the
curvature and the rescaling can be made locally but not globally. This
influences the dynamics of the fermions propagating in the texture in
the
manner similar to the gravitational field.

The question arises what and how big is the analog of the
gravitational
constant
$G$ in $^3$He-A and what is its temperature dependence. Since in
$^3$He-A the
analog of the Planck energy scale is played by gap amplitude
$\Delta(T)$ (see
Review
\cite{AxialAnomaly}), the gravitaional constant is of order
$G\sim 1/\Delta^2(T)$. However the quantitative evaluation of $G$ is
not
straghtforward:  because of the double role of the $\hat l$-field,
which
produces both the "electromagnetic" and "gravitational" effective
fields, it is
not easy to separate the   "electromagnetic" and "gravitational"
terms
in the
$^3$He-A Lagrangian.  The separation can be made only for specific
situations. For example,  in the discussion of the effect of axial
anomaly on the transformation of the fermionic charge into magnetic
field
and back, only the electromagnetic part of the $\hat l$ action was
involved \cite{AxialAnomaly}. Here we discuss the
opposite case when  the "electromagnetic" effects of the
$\hat l$-field are absent, and one has a pure "gravitational" field.
This
happens if the "magnetic" field
${\bf B}=\nabla
\times {\bf A}$ is absent. The radial disgyration (string) with $\hat
l=\hat
\rho$ and the point monopole with  $\hat l=\hat r$  are such textures,
since both have $\nabla
\times {\hat l}=0$. Here
$\hat z,
\hat \rho, \hat \phi$ and $\hat r, \hat \theta, \hat \phi$  are the
unit
vectors
of the cylindrical and spherical coordinate systems correspondingly.
Considering the energy of the radial disgyration and the metric
produced by
this topological singularity, one can deduce the effective
gravitational
constant
$G$, which couples these two quantities. Another estimation of $G$
follows from
the consideration of the so-called clapping mode of the order
parameter,
which is
the analog of graviton.

{\bf Conical singularity with negative angle
deficit.}
The radial disgyration is one of the topologically stable
linear defects in
$^3$He-A. This is an axisymmetric distribution of the ${\hat{\bf l}}$
vector
\begin{equation}
{\hat l}({\bf r})= {\hat\rho} ~~,
\label{RadialDisgiration}
\end{equation}
 with  the axis of the defect
line   along ${\hat z}$.
The interval corresponding to the metric in Eq.(\ref{gik}) is
\begin{equation}
ds^2=dt^2 - {1\over  c_\perp^2} dz^2 -{1\over   c_\parallel^2}
\left(dr^2 +  { c_\parallel^2 \over  c_\perp^2  } r^2 d\phi^2\right)
~~,
\label{IntervalForRDisg}
\end{equation}

Rescaling the radial and axial coordinates
 $ \rho=c_\parallel R$ , $z= c_\perp Z $
one obtains
\begin{equation}
ds^2=dt^2 - dZ^2- dR^2 -a^2 R^2  d\phi^2~~,
~~a^2=c_\parallel^2/ c_\perp^2>1.
\label{RadialDysgMetric}
\end{equation}
 In
relativistic theories such metric, but with $a^2<1$, arises outside
the local
strings.  The space outside the string core is
flat, but the proper length $2\pi R a$ of
the circumference of radius
$R$ around the axis is smaller than $2\pi R$, if $a<1$. In
our case we have $a^2>1$, i.e. the  "negative angle  deficit". The
conical
singularity gives rise to the curvarture which is concentrated at the
axis of
disgyration ($R=0$)
\cite{SokolovStarobinsky,Banados}:
\begin{equation}
  {\cal R}^{R\phi}_ {R\phi}  =2\pi{a-1
\over  a }\delta_2({\bf R}) ~~,~ ~ \delta_2({\bf R})=\delta (X)\delta
(Y).
\label{CurvatureRDysg}
\end{equation}

Such metric can arise from the Einstein equations for the local
cosmic string
with the singular energy density concentrated in the string core
\begin{equation}
 {\cal T}^{0}_ {0}  = {1-a
\over 4G a }\delta_2({\bf R})   ~~.
\label{NegativeEnergyDensity}
\end{equation}
where  $G$ is the gravitational constant. Since $a=c_\parallel /
c_\perp \gg
1$, this should be rather unusual cosmic string with a large negative
mass of
Planck scale.  If one finds such  singular  contribution to the
energy density of $^3$He-A in the presence of radial disgyration one
can identify the effective gravitational constant in  $^3$He-A.

Let us consider the distribution of the (orbital) order parameter --
the
complex
vector ${\bf e}$ -- in the radial disgyration:
\begin{equation}
{\bf e}({\bf r})=f(\rho)\hat \phi + i \hat z
~~,~~f(\rho=0)=0~~,~~f(\rho=\infty)=1  ~~.
\label{OrderParameter}
\end{equation}
This order parameter influences the energy spectrum of the
fermions in such a way, that it is equivalent to the effective metric
\begin{equation}
ds^2=dt^2 - dZ^2- dR^2 -a^2  f^{-2}(Rc_\parallel) R^2  d\phi^2~~ .
\label{GeneralRadialDysgMetric}
\end{equation}
The function $f(\rho)$ can be obtained from the Ginzburg-Landau free
energy
functional, Eq.(5.4) + Eq.(7.17) in \cite{Vollhardt1990}, which for
the chosen
Anzats Eq.(\ref{OrderParameter}) has the form
\begin{equation}
F=K{v_Fk_F^2\over 96\pi^2}  \int_0^{z_0} dz \left\{
\int_{\rho<\rho_0} d^2\rho
\left[
\Lambda (1-f^2)^2 + {f^2\over
\rho^2} +
\left({df\over d\rho}\right)^2 \right]-2\pi\int_0^{\rho_0} d\rho
{df^2\over
d\rho}\right\} ~~.
\label{FreeEnergy}
\end{equation}
Here $\rho_0$ and $z_0$ are the radius and the height of the
cylindrical
vessel with the disgyration on the axis;   $\Lambda \sim
\Delta^2/v_F^2$; the overall dimensionless factor $K$ in the Ginzburg-
Landau
region close to the transition temperature $T_c$ is
\begin{equation}
 K(T) = 1-{T^2\over T_c^2} ~~,~~T\rightarrow T_c~.
\label{K(Tc)}
\end{equation}

 One can see that the first term
in the curly brackets in Eq.(\ref{FreeEnergy}) is some kind of the
dilaton
field,
while the second term is just what we need: it is the pure divergence
and thus
can be represented as the singular term, which does not depend on the
exact
structure of the disgyration core, but nevertheless contributes the
core
energy:
\begin{equation}
{\cal F}_{\rm div}= -2\pi K {v_Fk_F^2 \over 96 \pi^2}  \delta_2({\bf
\rho})
~~,~~F_{\rm div}=\int d^3x {\cal F}_{\rm div}= -  K{v_Fk_F^2 \over 48
\pi}
  ~~.
\label{SingularFreeEnergy}
\end{equation}
Now let us extract the constant $G$ by comparing this core energy
with the
string mass
$M$  obtained by integration of ${\cal T}^0_0$:
\begin{equation}
 M= \int d^3X\sqrt{-g} {\cal T}^{0}_ {0}  =  {1-a
\over 4G} Z_0    ~~.
\label{NegativeMass1}
\end{equation}
Translating this to the $^3$He-A language, where the "proper" length
is
$Z_0=z_0
/c_\perp$,  and taking into account
that $a =c_\parallel/c_\perp \gg 1$ one has
\begin{equation}
 M=-   {c_\parallel
\over 4G c_\perp^2} z_0 ~~.
\label{NegativeMass2}
\end{equation}
Then from equation, $F_{\rm div}=M$, one obtains the gravitational
constant \begin{equation} G(T)=  {12 \pi \over  K(T)\Delta^2(T)}~.
\label{GravitationConstant}
\end{equation}

Though we cannot extrapolate the temperature dependence of $K(T)$ in
Eq.(\ref{K(Tc)}) to the low
$T$, we can expect that the overall temperature dependence can be
approximated
by
\begin{equation}
G(T)\sim G(0)  \left(1-{T^2\over T^2_c}\right)^{-2}~,~ G(0)\sim
{1\over
T_c^2} \sim {1\over
\Delta^2(0)}~,
\label{OverallTemperatureDependence}
\end{equation}
where $G(0)$ is the value of $G$ at $T=0$.

The negative mass $M$ does not mean that the vacuum is unstable
towards
formation of the string: the  energy of the
radial disgyration is dominated by the first (dilatonic) term in the
curly
brackets in Eq.(\ref{FreeEnergy})
\begin{equation}
 E_{\rm disg}(\rho_0)=K{v_Fk_F^2\over 48\pi } z_0\ln {\rho_0 \Delta
\over
c_\parallel}
  ~~.
\label{DisgyrationEnergy}
\end{equation}
Translating this to the relativistic language one obtains
\begin{equation}
 E_{\rm disg}(R_0)= {a\over 4G } Z_0\ln {R_0   \over
R_{\rm Planck}}
  ~~,
\label{DisgyrationEnergyRel}
\end{equation}
where $R_{\rm Planck}=1/\Delta$ is the Planck radius.
This energy is larger by the logarithmical
factor than the negative "mass of matter" $F_{\rm div}=M$ in
Eq.(\ref{NegativeMass1}) related to the string core. So the situation
in this
example of the $^3$He-A gravity is as follows: The energy in
Eq.(\ref{DisgyrationEnergy}) is not gravitating, but it determines
the metric.
This metric through the Einstein equation gives rise to the negative
mass $M$,
which then contributes to the core energy of disgyration.   This is
an example
of how in the effective theory the vacuum is not gravitating but
determines the
metric.

{\bf Gravitational constant from graviton energy-momentum.}
An independent estimation of $G$ in $^3$He-A is obtained using the
energy
 of the graviton field. In the relativistic theory the energy
density of the graviton propagating along $Z$ is
\begin{equation}
 {\cal T}^0_0=   {1   \over
16 \pi G} \left[(\partial_Z h_{XY})^2 + {1\over 4}((\partial_Z
(h_{XX} -h_{YY}))^2 \right]  ~~.
\label{Graviton EnergyDensity}
\end{equation}

Let us consider the corresponding energy density in $^3$He-A. For this
purpose we
choose the complex order parameter vector ${\bf e}$ in the form
\begin{equation}
{\bf e}(z)=\left[\left( 1+ {1\over 2} h_{XX}(z)\right) \hat x
+{1\over 2}
h_{XY}(z)
\hat y\right]+ i\left[\left( 1+ {1\over 2} h_{YY}(z)\right) \hat y  +
{1\over 2}
h_{XY}(z) \hat x\right] ~~.
\label{GravitonOrderParameter}
\end{equation}
It corresponds to the following effective metric for the
quasiparticles
propagating on the background of this order parameter
\begin{equation}
ds^2=dt^2 - dZ^2- (1+h_{XX}(Z))dX^2 -(1+h_{YY}(Z))dY^2 -
2h_{XY}(Z)dXdY~~ .
\label{GravitonMetric}
\end{equation}
The gradient part of the Ginzburg-Landau free energy
functional
for this
Anzats, Eq.(\ref{GravitonOrderParameter}), has the form
\begin{eqnarray}
{\cal F}=K{v_Fk_F^2\over 96\pi^2} \left(\nabla_i {\bf e}\nabla_i {\bf
e}^*+
\nabla_i e_j\nabla_j e_i^*+ \nabla_i e_i \nabla_j e_j^*\right)
\label{GradientEnergy}
\\ =K{v_Fk_F^2\over 192\pi^2} \left[ {1\over
4}((\partial_z (h_{XX} + h_{YY}))^2 + (\partial_Z h_{XY})^2+{1\over
4}((\partial_z (h_{XX} -h_{YY}))^2   \right] ~.
\label{ClappingModeEnergy}
\end{eqnarray}
The first term in Eq.(\ref{ClappingModeEnergy}) describes the so-
called
pair-breaking mode of the order parameter, which is the analog of the
(spin 0)
dilaton energy. The other two terms describe  the so-called clapping
mode,
which
corresponds to the (spin 2) graviton
\cite{Exotic}. From the comparison of the clapping mode energy in
Eq.(\ref{ClappingModeEnergy}) and the graviton energy in
Eq.(\ref{Graviton
EnergyDensity}), one obtains the same value for the gravitational
constant as in Eq.(\ref{GravitationConstant}).

{\bf Monopole.}
In $^3$He-A the monopole is the hedgehog in the $\hat l$-field, $\hat
l=\hat
r$, which is the termination point of the quantized vortex (the
vortex is analogous to the spinning or torsion string).  The
effective metric
far from the string is \begin{equation} ds^2= \left(dt +{\hbar\over
2m_3c_\perp^2}(1-\cos\theta)d\phi\right)^2 -{1\over
c_\parallel^2}dr^2 -
{1\over c_\perp^2}r^2 (d\theta^2+\sin^2\theta d\phi^2)~~.
\label{MonopoleMetric1}
\end{equation}
The spinning string terminating on the monopole can be removed by the
following
trick. Let us introduce the electrically charged $^3$He-A, i.e.
the superconductor with the $^3$He-A order parameter. Then put the
t'Hooft-Polyakov magnetic monopole  to the hedgehog. In this
case the Abrikosov string will be cancelled and thus one obtains only
the point
singularity in the $\hat l$-field -- the hedgehog -- with pinned
magnetic
monopole
\cite{AphaseMonopole}.  After rescaling of the radial coordinate
$R=r/ c_\parallel$
one obtains the metric
\begin{equation}
ds^2=dt^2 - dR^2 -a^2 R^2(d\theta^2+\sin^2\theta d\phi^2)~~,
~~a^2=c_\parallel^2/ c_\perp^2 .
\label{GlobalMonopoleMetric}
\end{equation}
This metric
describes the 3D conical singularity \cite{SokolovStarobinsky}.  In
relativistic theories such metric, but with $a^2<1$, arises for the
global
monopoles \cite{GlobalMonopole1,GlobalMonopole2}. In our case we have
$a^2>1$, i.e. the "negative   deficit" of the solid angle. Such
situation is
also different from the solid angle deficit $>4\pi$ discussed in
\cite{ChoVilenkin}, where $a^2<0$ giving rise to instability of the
stationary monopole and to the inflation.
 The nonzero curvarture
elements and the corresponding energy density of matter are
\cite{SokolovStarobinsky}
\begin{equation}
 {\cal R}^{\theta\phi}_ {\theta\phi}  =-{1-a^2
\over  a^2  R^2}~,~{\cal R}=2{\cal R}^{\theta\phi}_ {\theta\phi} ~~,~~
 {\cal T}^{0}_ {0}  = {1-a^2
\over  a^2 8\pi G R^2}  ~~.
\label{NegativeEnergyDensityMonopole}
\end{equation}
Integrating the energy density one obtains the negative
contribution to
the monopole energy (at $a\gg 1$):
\begin{equation}
M(R_0)=\int_{R<R_0} d^3R \sqrt{-g} {\cal T}^{0}_ {0}  = -  {a^2
\over  2 G  } R_0
\label{NegativeMonopolMass}
\end{equation}
Translating to the $^3$He-A language with the value of $G$ from
Eq.(\ref{GravitationConstant}) and with $r_0=R_0c_\parallel$  being
the radius
of spherical vessel one obtains
\begin{equation}
M(r_0)= -  {c_\parallel
\over  2 c_\perp^2 G  } r_0 =- { K (T)\over 24\pi} v_Fk_F^2 r_0 ~.
\label{NegativeHedgehogMass}
\end{equation}
The real energy of the radial $\hat l$-texture (without the attached
spinning string) is
$2|M(r_0)|$ as follows from  Eq.(\ref{GradientEnergy}). However we
cannot
unambiguously identify the obtained negative mass with  some specific
term in
$^3$He-A: all the terms have the same
$1/r^2$-dependence. This is distinct from the case of disgyration,
where  the negative contribution has the
$\delta$-function singularity and we could easily identify it with
the similar
negative-energy
$\delta$-function contribution in the
$^3$He-A action.  One possibility, that this negative mass can be
identified
as the  interaction of the magnetic field of the monopole with the
orbital
momentum of $\hat l$-field in the charged $^3$He-A, will be discussed
elsewhere.

{\bf Conclusion.}
We considered two metrics with nonzero curvature arising in the
$^3$He-A
textures. The same metrics occur  outside the  local cosmic string
and the
global cosmic monopole, both with the negative mass. In the case of
the local
cosmic string the negative energy comes from the $\delta$-function
singularity of the curvarture. For the
$^3$He-A disgyration the negative energy contribution to the textural
energy
also comes from
$\delta$-function term in the action. Identifying these two negative
energy
terms, we obtained the  value of
the effective gravitational constant
$G(T)$ in $^3$He-A.  The same value is obtained from the energy-
momentum
tensor for the analog of the graviton in $^3$He-A. $G(T)$ is inversely
proportional to the square of the "Planck" energy and depends on
$T$ increasing with $T$, which corresponds to the vacuum screening of
the
gravity. The temperature dependence of the gravitational
constant leads to its time dependence during the evolution of the
Universe.
The latter has been heavily discussed starting with the Dirac
proposal (see
Review\cite{Barrow}).

>From the $^3$He-A consideration it appeared that there are two
contributions to
the temperature dependence of
$G$ in Eq.(\ref{GravitationConstant}).  (i) The dependence, which
comes
from the
factor $K(T)$, is the traditional one. Since the effective gravity is
obtained
from the integration over the fermionic (or bosonic) degrees of
freedom
\cite{Sakharov}, it is influenced by the thermal distribution of
fermions. In
relativistic theories, even
at low $T$ the renormalization of $G$ is model-dependent: it depends
not
only  on the fermionic and bosonic fields, but also on the cut-off
function
if the gravitons are included\cite{Reuter}. (ii) Another source of the
temperature dependence of $G(T)$ in Eq.(\ref{GravitationConstant}) is
that
the "Planck" energy cut-off $\Delta(T)$ depends on temperature. Even
at low
$T$ this dependence is determined by the transPlanckian physics.

I thank Matt Visser, who pointed out that the metric induced by the
radial disgyration in $^3$He-A corresponds to the negative mass of the
string in
the relativistic theory, and Alexei Starobinsky for illuminating
discussions.

\end{document}